\documentclass[fleqn,usenatbib]{mnras}

\usepackage{newtxtext,newtxmath}

\usepackage[T1]{fontenc}

\DeclareRobustCommand{\VAN}[3]{#2}
\let\VANthebibliography\thebibliography
\def\thebibliography{\DeclareRobustCommand{\VAN}[3]{##3}\VANthebibliography}

\usepackage{graphicx}	
\usepackage{amsmath}	
\usepackage{xcolor}

\title[How bright can old magnetars be?]{How bright can old magnetars be? Assessing the impact of magnetized envelopes and field topology on neutron star cooling}

\author[C. Dehman et al.]{
Clara Dehman,$^{1,2}$\thanks{E-mail: c.dehman@csic.es}
José A. Pons,$^{3}$ Daniele Viganò,$^{1,2,4}$  Nanda Rea$^{1,2}$ \\
$^{1}$Institute of Space Sciences (ICE-CSIC), Campus UAB, Carrer de Can Magrans s/n, 08193, Barcelona, Spain\\
$^{2}$Institut d'Estudis Espacials de Catalunya (IEEC), Carrer Gran Capità 2–4, 08034 Barcelona, Spain\\
$^{3}$Departament de Física Aplicada, Universitat d'Alacant, 03690 Alicante, Spain \\
$^{4}$Institute of Applied Computing \& Community Code (IAC3), University of the Balearic Islands, Palma, 07122, Spain\\
 }

\date{Accepted 2023 January 4. Received 2022 December 23; in original form 2022 November 4}

\pubyear{2022}

\begin{document}
\label{firstpage}
\pagerange{\pageref{firstpage}--\pageref{lastpage}}
\maketitle

\begin{abstract}
Neutron stars cool down during their lifetime through the combination of neutrino emission from the interior and photon cooling from the surface. Strongly magnetised neutron stars, 
called magnetars, are no exception, but the effect of their strong fields adds further complexities to the cooling theory.
Besides other factors, modelling the outermost hundred meters (the envelope) plays a crucial role in predicting their surface temperatures. 
In this letter, we revisit the influence of envelopes on the cooling properties of neutron stars, with special focus on the critical effects of the magnetic field.
We explore how our understanding of the relation between the internal and surface temperatures has evolved over the past two decades, 
and how different assumptions about the neutron star envelope and field topology lead to radically different conclusions on the surface temperature and its cooling with age. 
In particular, we find that relatively old magnetars with core-threading magnetic fields are actually much cooler than a 
rotation-powered pulsar of the same age. This is at variance with what is typically observed in crustal-confined models. Our results have important implications for the estimates of the X-ray luminosities of aged magnetars, and the subsequent population study of the different neutron star classes.

\end{abstract}



\section{Introduction}
\label{sec: introduction}

It has long been hoped that observations of direct thermal emission from the surface of neutron stars (NSs), confronted to theoretical cooling curves (the temperature-age or luminosity-age relation), could yield valuable information about
star interior, such as the nuclear equation of state and chemical composition \citep{yakovlev2004,page2006,potekhin2015,pons2019}. However, NSs are also known to be endowed with strong magnetic fields, and therefore an appropriate treatment of the coupled thermal and magnetic field evolution in detail is of great importance to understand the observed emissions from the surface of NSs. In this respect, recent works have devoted a significant effort to extend realistic simulations to 3D \citep{wood2015,gourgouliatos2016,grandis2020,degrandis21,igoshev21a,igoshev21,dehman2022}.

Within this context, the relevance of envelope models is usually overlooked, since it enters in multidimensional simulations "only" as a boundary condition. However, as we will show, it actually plays a key role to connect the internal properties with the observable quantities (effective temperature, luminosity), especially for highly magnetised objects, such as magnetars. 

The very different thermal relaxation timescales of the envelope and the crust of NSs make computationally unfeasible any attempt to perform cooling simulations in a numerical grid that includes all layers up to the star surface. Thus, the usual approach is to compute envelope models separately, and then use a phenomenological fit predicting the value of the local surface temperature ($T_s$) as a function of the temperature at the base of the envelope ($T_b$), to be used as a boundary condition (the $T_b-T_s$ relation). 

%
\begin{figure*}
\includegraphics
 [width=.43\textwidth]
{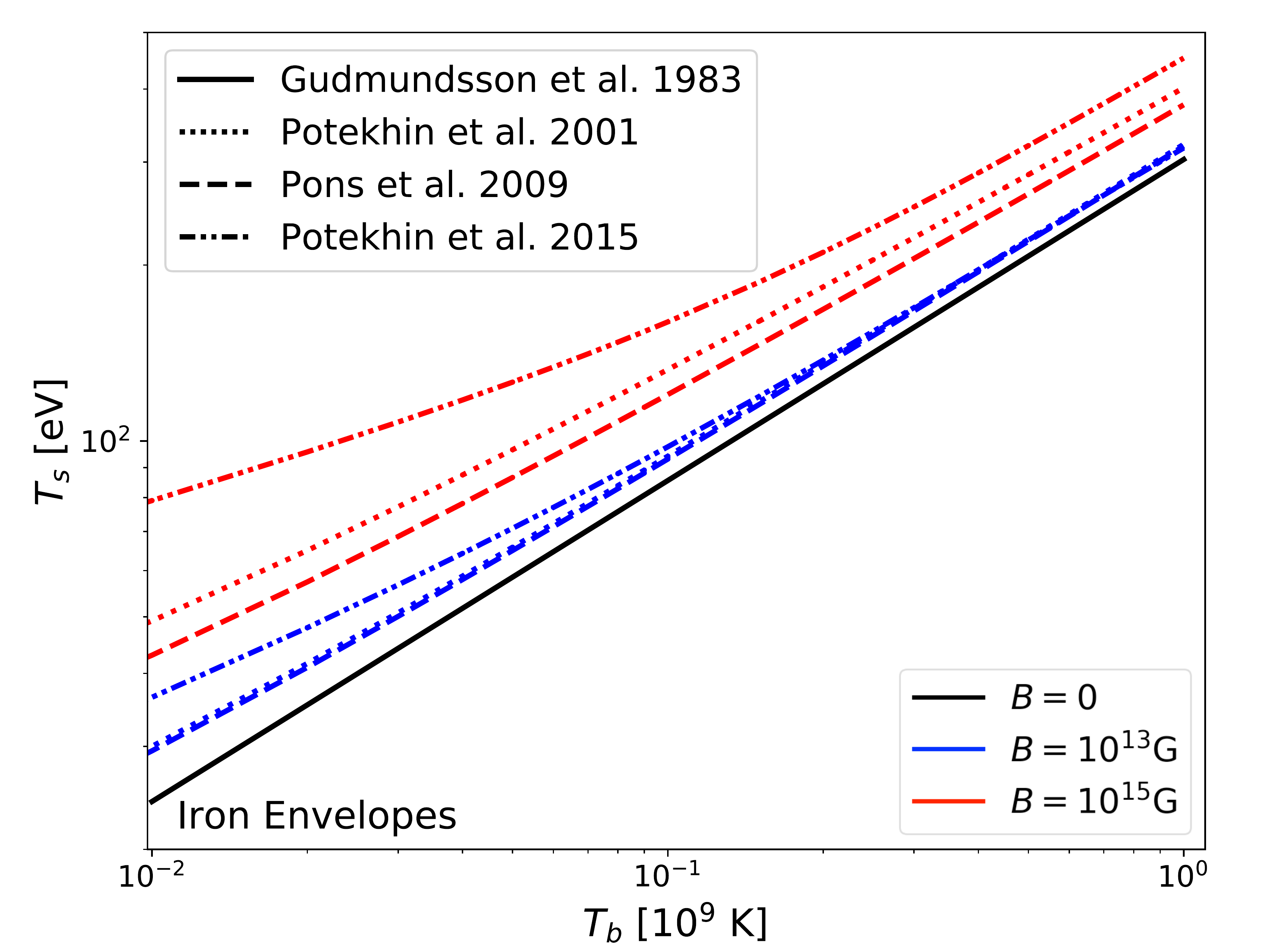}
\includegraphics
[width=.43\textwidth]
{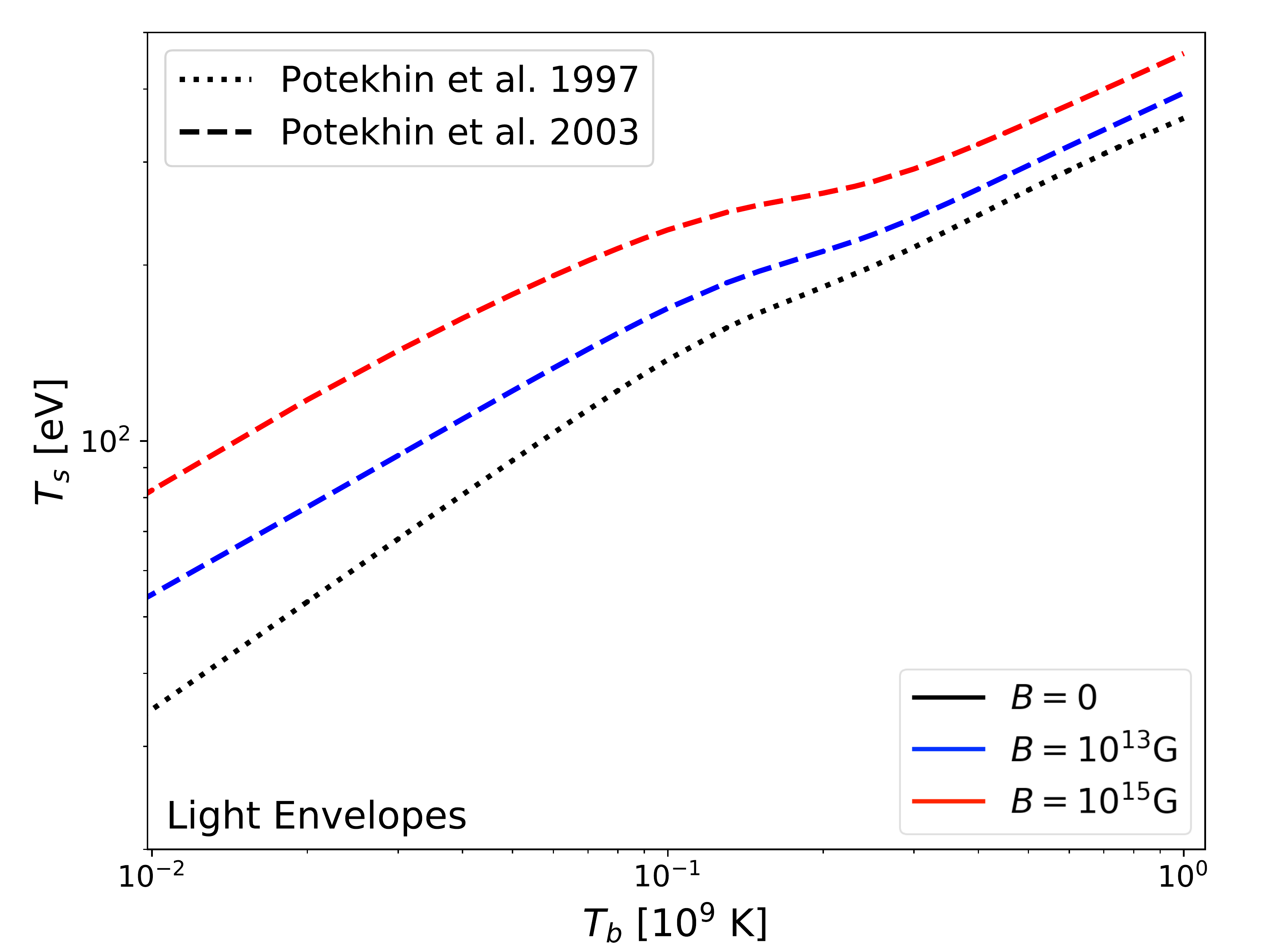}
\caption{$T_b-T_s$ relations of different envelope models. The non-accreted models are illustrated on the left and the fully accreted ones in the panel on the right. The studied envelopes are: (i) non-magnetised (in black) \citep{gudmundsson1983,potekhin1997}, and (ii) magnetised (in color) \citep{potekhin2001,potekhin2003,pons2009,potekhin2015}. For the latter, we consider two different values of a purely radial magnetic field strength (then, suitable for a polar $T_s$ if the topology is a simple dipole): $B=10^{13}$ (in blue) and $10^{15}$ G (in red).}
\label{fig: envelope models Tb-Ts}
\end{figure*} 


Among the many early studies of the thermal structure of NSs, we must mention the seminal works of \cite{tsuruta1971}, \cite{gudmundsson1983}, 
\cite{hernquist1985} or \cite{schaaf1990}, who pointed out that regions with tangential magnetic field are much colder than the regions where the field is nearly radial (see \cite{yakovlev1994} and references therein for a review of the early works).
Later, \cite{potekhin1997} constructed a more general fit valid for different compositions. These envelope models used improved calculations on the equation of state and opacities in the outer NS layers. In particular, the so-called accreted envelopes contain layers of different chemical elements (H, He, C, O shells) created from accreted matter from the supernova fallback material.

 \cite{page1994} and \cite{page1996} were the first to describe realistic surface temperature distributions with dipolar and dipolar+quadrupolar magnetic fields, the latter presenting "$T_b - T_s$" relationships with such configurations. The thermal structure of NSs with magnetized envelopes was also studied by \cite{potekhin2001}, and later improved in \cite{potekhin2003}. It includes the effect of magnetic fields on the $T_b-T_s$ relation, providing analytical fits valid for a magnetic field strength up to $10^{16}$ G and arbitrary inclination angles of the field lines with respect to the normal to the surface. Similar studies
 exploring other field topologies were done by \cite{geppert2004,geppert2006} and \cite{perezazorin2006}.
Subsequent calculations in \cite{potekhin2007} included the effect of the neutrino emissivity in the outer crust. \cite{pons2009} revisited the magnetized envelope problem with two motivations: (i) upgrading the microphysical inputs (thermal conductivity) because the contribution of ions or phonons to the thermal conductivity of the envelope can reduce the anisotropy of heat conduction \citep{chugunov2007}; (ii) estimating the accuracy of the plane-parallel approximation since its spherical symmetry assumption does not allow meridional heat fluxes.

The state-of-the-art models can be found in the thorough review by \cite{potekhin2015}. They present new fits for non-accreted magnetised envelopes, including both the effects of neutrino emission and the effects of non-radial heat transport.

In this paper, we aim at comparing a set of the different envelope models studied in the literature.
Our main goal is to assess how the evolution of theoretical cooling models for different magnetic fields intensities and geometries are affected by the choice of the envelope and its treatment. In light of this analysis, to determine under which circumstances we can use observational X-ray data to constrain the cooling models, and consequently the NS parameters.

The letter is organized as follows. In \S\ref{sec: Envelope models}, we recap the $T_b-T_s$ relation of the different envelope models existing in the literature with different magnetic field intensities. In \S\ref{sec: cooling models}, we perform cooling simulations using the last version of our 2D magneto-thermal code \citep{vigano2021}. We examine crustal-confined and core-dominant field topologies considering both iron and fully-accreted light envelopes. We discuss our results and draw our main conclusions in \S\ref{sec: conclusions}. 

\vspace{-5mm}

\section{Envelope models}
\label{sec: Envelope models}

Since NSs are observed both as isolated sources or as part of binary systems, it is common to consider two different compositions for the envelope: either iron, arguably expected in the case of catalyzed matter in isolated systems, or light elements, mainly thought as products of accretion from a companion star or in a newly born systems that witness fall-back accretion after a supernova. In this study, we explore four models composed of iron (non-accreted matter) and two models of fully accreted (light) envelopes \citep{gudmundsson1983,potekhin1997,potekhin2001,potekhin2003,aguilera2008,pons2009, potekhin2015}.

In essence, an envelope model is simply a stationary solution for the heat transfer equation and it is then fitted to give an empirical relation between the surface temperature $T_s$, which determines the radiation flux, and the interior temperature $T_b$ at the crust/envelope boundary. The location of $T_b$ is generally chosen to correspond to some density between the neutron drip point $\rho = 3 \times 10^{11}$ g cm$^{-3}$ and $\rho = 10^{10}$ g cm$^{-3}$. At such a low density, the neutrino emission is usually negligible, as long as $T_b < 10^9$ K (which happens very soon, only a few decades after the NS birth). Thus, we have omitted corrections due to neutrino emissivity. 

The $T_b-T_s$ relations of the studied envelope models are illustrated in Fig. \ref{fig: envelope models Tb-Ts}. In the panel on the left we show the iron envelopes, whereas the light ones are displayed on the right. The studied envelopes are: (i) two models with no magnetic field dependence (in black), e.g., \cite{gudmundsson1983} in the left panel (solid lines) and \cite{potekhin1997} in the panel on the right (dots);
(ii) other magnetised envelopes, for which we show the $T_s$ for two values of a purely radial surface magnetic field strength (i.e., suitable for a pole in a dipolar topology): $B=10^{13}$ G (in blue) and $B=10^{15}$ G (in red).
\cite{potekhin2001} is illustrated with dots (left panel), 
\cite{potekhin2003} with dashed lines (right panel), \cite{pons2009} with dashed lines (left panel), and finally \cite{potekhin2015} with dashdotdotted lines (left panel). 

\begin{figure*}
\includegraphics[width=.8\textwidth]
{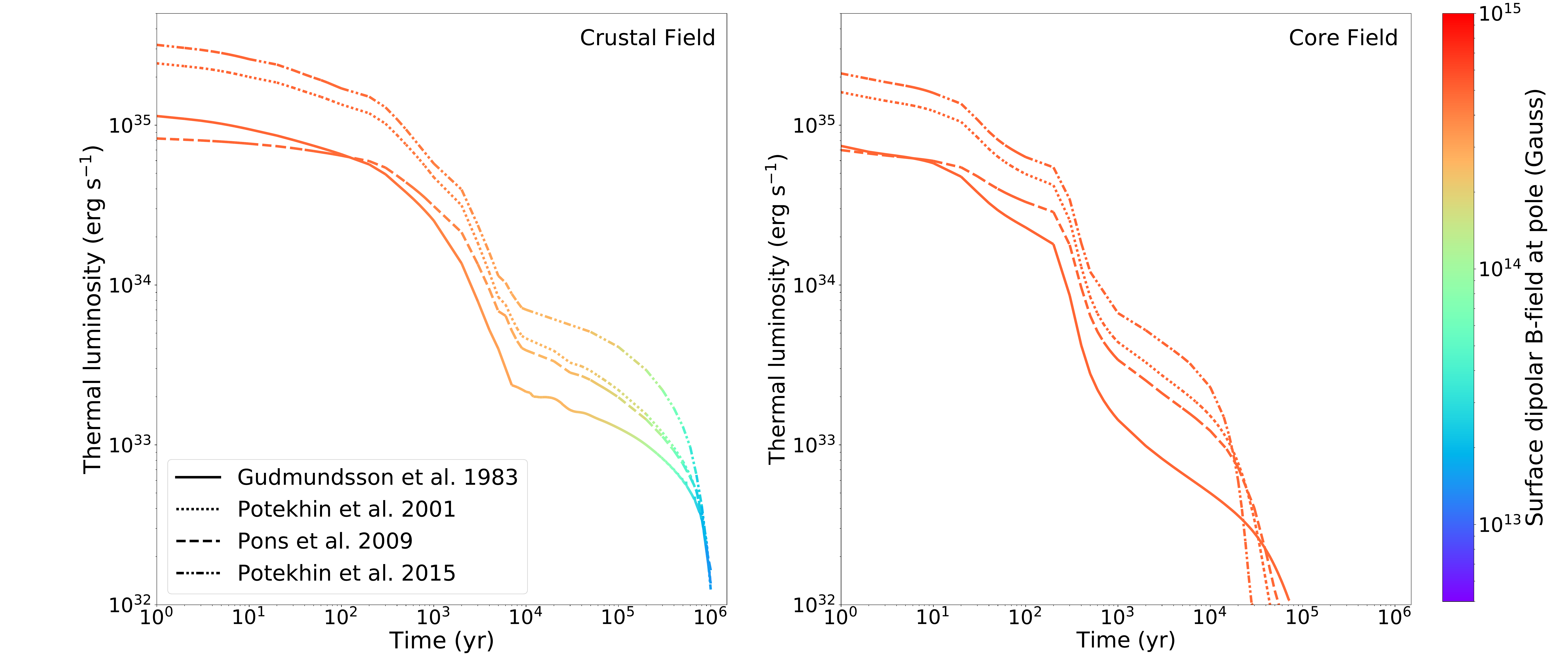}
\caption{Luminosity curves of the four studied iron envelopes \citep{gudmundsson1983,potekhin2001,pons2009,potekhin2015} with an initial magnetic field intensity at the polar surface of $B=5 \times 10^{14}$ G (hereafter, the colorbar indicates its evolution). The left panel corresponds to crust-confined topology, whereas the right one to core-dominant field topology.}
\label{fig: cooling curves}
\end{figure*} 


The lowest effective temperature $T_s$ among all models is displayed by \cite{gudmundsson1983}. Every new effect incorporated in later works (composition, magnetic field) results in a higher predicted surface temperatures $T_s$ for a given $T_b$. Let us briefly review the main conclusions from a quick comparison of models.

In general, it is well known that assuming light-elements envelopes appreciably affect the NS luminosity \citep{potekhin1997}. 
Compared to iron models, we have a higher $T_s$ for the same $T_b$.
Concerning magnetic fields, as long as the average intensity is $B \lesssim 10^{13}$ G, we expect a surface temperature similar to the non-magnetised case (for the same given composition).
On the other hand, for magnetar conditions, the general trend is that higher fields lead to have higher surface temperature, everything else being equal. Interestingly, the left panel of Fig. \ref{fig: envelope models Tb-Ts} shows that the more recent calculations (incorporating more accurate physics) have revised the predicted $T_s$ to higher values than any of the previous works.

Thus, it is expected that the state-of-the-art envelope models predict different cooling curves from the models used two decades ago. This motivates us to revisit the results for cooling curves and consider different magnetic field topologies and strengths, as an important step in understanding the observational data. 

\vspace{-5mm}

\section{Neutron star cooling models}
\label{sec: cooling models}

The cooling history of a magnetar is a delicate balance between neutrino and photon emissivity on one side and Joule heating in the star's crust on the other. If the currents are dissipated in the outer crust, the heat deposited is more effectively transported to the surface and has an impact on the star luminosity. On the contrary, heat dissipated in the inner crust or the core is very inefficient in modifying the surface temperature, because it is essentially lost via neutrino emission, as first discussed in \cite{kaminker2006} to explain the high thermal luminosities of magnetars.

To compare the different envelopes existing in the literature,
we have used the 2D magneto-thermal code (the latest version is described in \citealt{vigano2021}) to run a set of cooling models using different initial configurations.
The NS background model is a $1.4 M_\odot$ NS built with the Sly4\footnote{\url{https://compose.obspm.fr/}} equation of state \citep{douchin2001}, and we assume the superfluid models of \cite{ho2015}, which is the reason for the abrupt change in the slope of the cooling curves at ages $\sim 300$ yrs in, e.g., the right panel of Fig. \ref{fig: cooling curves}.
The rapid cooling during the photon cooling era is also caused by the low core heat capacity, which in turn depends on the assumed pairing details. 
A comprehensive revision of the microphysics embedded in magneto-thermal models can be found in \cite{potekhin2015}.

We considered two families of magnetic field topologies to study in detail the two extreme configurations: (i) crust-confined field consisting of a poloidal dipole and a toroidal quadrupole with steep radial gradients;
(ii) core-dominated twisted-torus magnetic fields as in \cite{akgun17}, i.e., a dipolar topology, with the currents circulating almost only in the core, and Gyr-long decay timescales. We stress that, for our purposes, we choose these two extreme topologies mean to cover a wide range of values for the crustal Ohmic dissipation.
For each topology, we consider two different field strengths ($10^{13}$ G and $5\times 10^{14}$ G) for the initial value of the dipolar field at the polar surface. The maximum initial toroidal field is fixed to $10^{13}$ G in all cases. The magnetic field at the surface is always matched continuously with a current-free magnetic field (i.e. the electric currents do not leak into the magnetosphere $\boldsymbol{\nabla} \times \boldsymbol{B}= 0$, with vanishing field at infinity).

We have used the different envelope models presented in \S\ref{sec: Envelope models} coupled with the NS cooling models, to study the dependence of the NS cooling curves on the assumed envelope, in two given magnetic field topologies. Magnetar cooling curves obtained using different iron envelopes and a field strength of $B=5 \times 10^{14}$ G are shown in Fig. \ref{fig: cooling curves}. In the left panel, we consider a crustal-confined topology, and in the right panel we have a core-dominant field.
For a high field intensity, e.g., magnetar-like scenario, there are significant qualitative differences between crustal-confined and core-dominant field. Let us summarize the main findings:
\begin{itemize}
    \item At early times, during the neutrino cooling era (say $t < 10^4$ yr), both models are similar. The interior temperature evolves independently of the envelope model (photon radiation is negligible), and the different $T_b-T_s$ relation translates directly in the surface temperature. Interestingly, the most recent models show the highest luminosities. This is a direct consequence of the results of Fig. \ref{fig: envelope models Tb-Ts} (left panel).
    \item Later, once we enter the photon cooling era, the situation is inverted. The envelope models that provide a higher surface temperature actually radiate photons (which now govern the evolution) more efficiently, and the star cools down faster.
    \item In this epoch, the difference between crustal-confined and core-threading magnetic field becomes more evident. In the first case, heat dissipation occurs relatively close to the surface, which keeps the stellar crust warmer and delays the drop of the luminosity. In the second case, Joule heating is completely inefficient (currents are mostly in the core), and the effect mentioned above, with a very fast drop of luminosity for high field models becomes evident.
\end{itemize}

\begin{figure}
\includegraphics
 [width=.43\textwidth]
{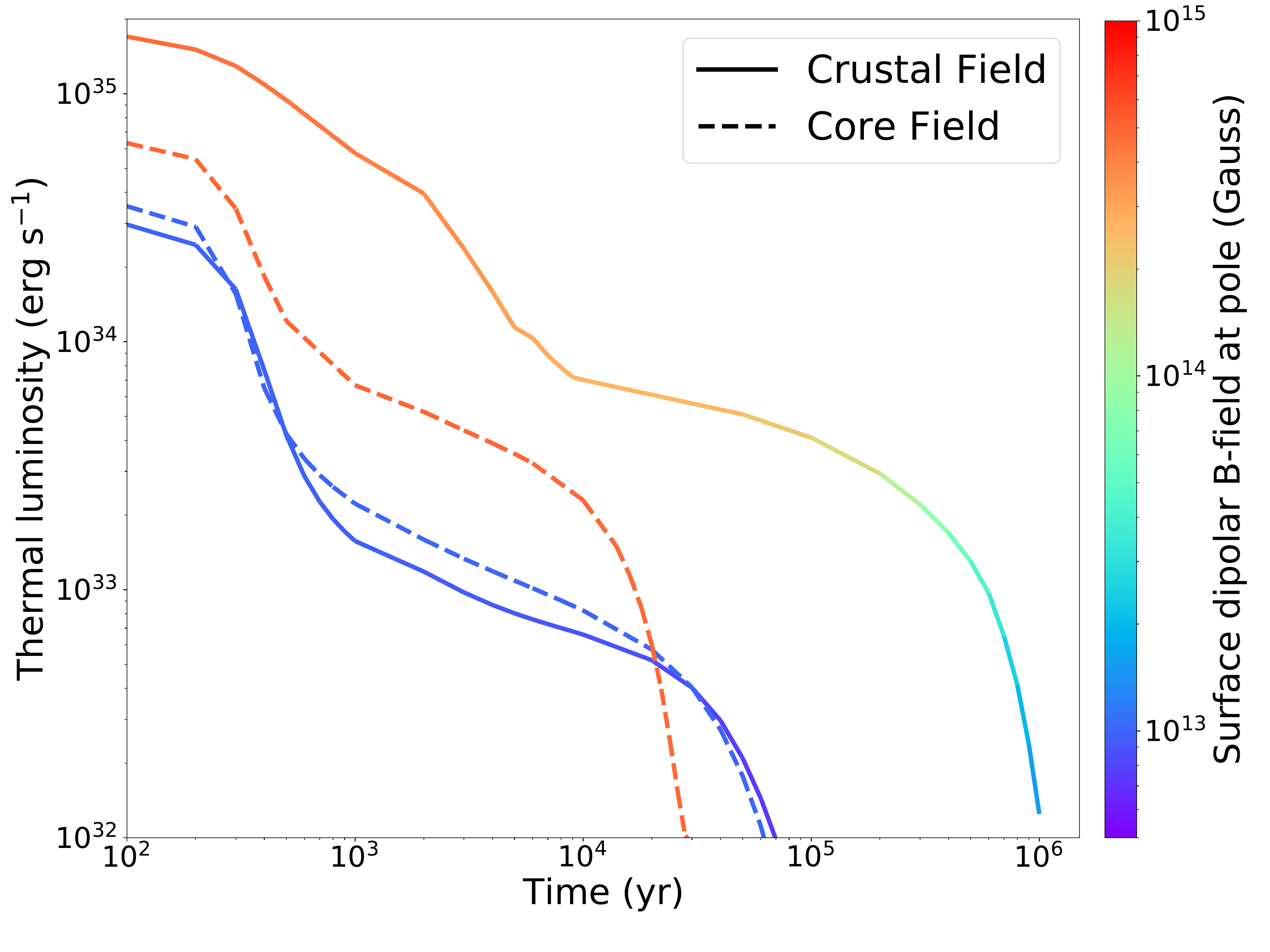}
\caption{Luminosity curves of the latest magnetised iron envelope \citep{potekhin2015} with different initial magnetic field intensities. Solid lines correspond to crustal field models and dashed-curves to core field ones.}
\label{fig: potekhin et al. 2015 cooling curves}
\end{figure}

To illustrate more clearly these differences, in Fig. \ref{fig: potekhin et al. 2015 cooling curves} we compare
cooling curves adopting the \cite{potekhin2015} envelope but now varying both, the field topology and strength. 
We show the results with the two different magnetic field intensities. For a relatively low magnetic field, e.g., $10^{13}$ G, the crustal-confined and core-dominated simulations have a very similar behavior and the magnetic field does not dissipate much (the curves keep the blue color throughout the evolution). For the strong field case, $5 \times 10^{14}$ G, the crustal-confined models show a significant dissipation of the magnetic field, e.g., the magnetic field has dissipated from $5\times 10^{14}$ G (red) to a few $10^{13}$ G (turquoise) after 1 million year of evolution (colorbar of Fig. \ref{fig: potekhin et al. 2015 cooling curves}). As a consequence, the impact of Joule heating is essential in the crust-confined, while it is almost negligible for the core-dominated model here considered, since the crustal currents are orders of magnitude less intense and the core currents have much longer Ohmic timescales (moreover, the little they dissipate converts into neutrinos). The most relevant difference is that core-threaded field simulations with magnetized envelopes show faster cooling after $10^4$ yr than low field models.
Therefore, the observational appearance of a magnetar at late times essentially depends on where currents are located and how much magnetic flux penetrates the core.

We also note that, for a strong enough magnetic field in the core of a NS, e.g., $B= 10^{15}$ G, an additional cooling channel via neutrino synchrotron \citep{kaminker1997} is activated. It provides further cooling of the NS. However, we found that this effect is subdominant.  

\begin{figure*}
 \includegraphics[width=16cm, height=9.2cm]
 {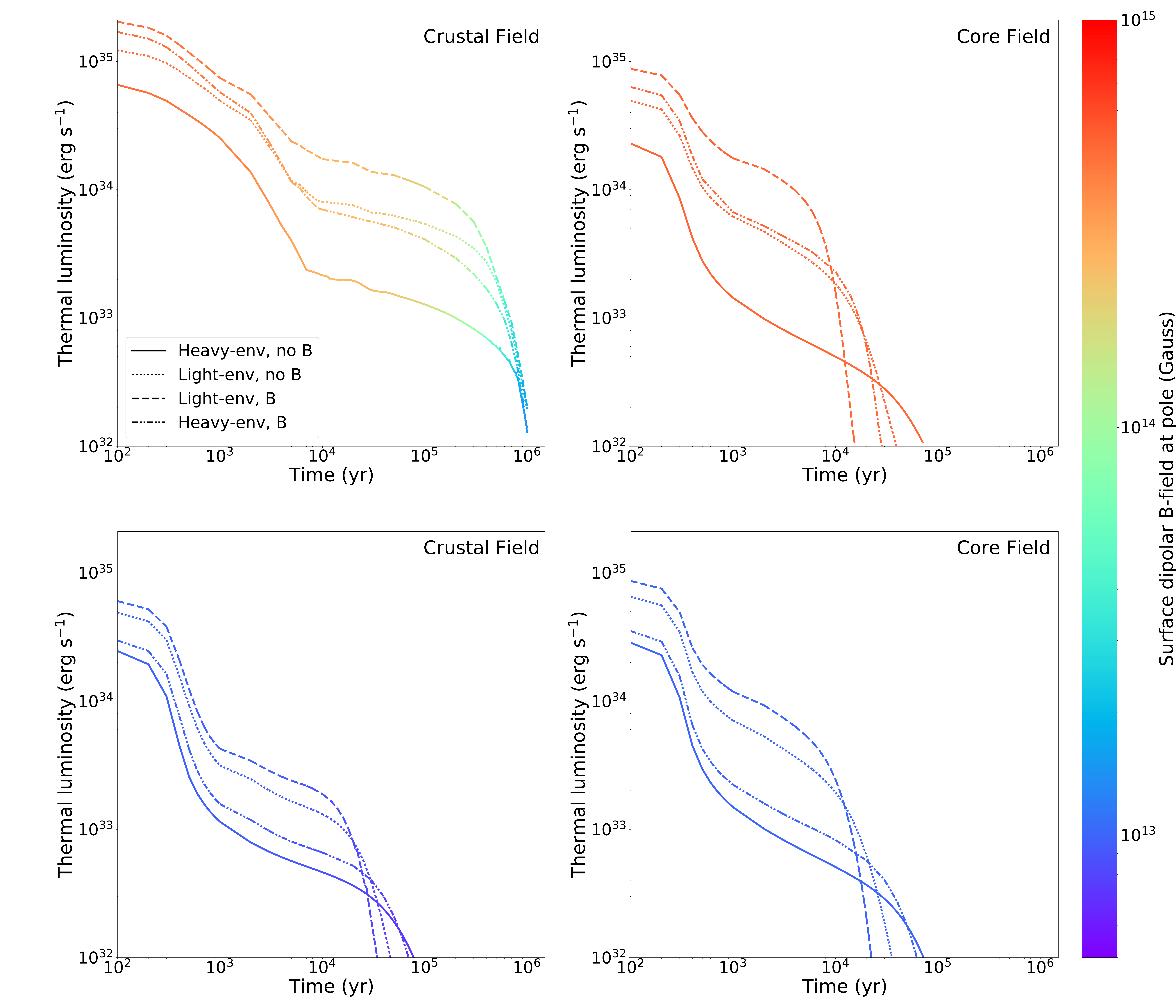}
\caption{Luminosity curves of four studied envelope models: Heavy-env, no B \citep{gudmundsson1983}, Light-env, no B \citep{potekhin1997}, 
Light-env, B \citep{potekhin2003} and Heavy-env, B \citep{potekhin2015}. On the left-hand side, we show the results of models with crustal-confined magnetic field, and on the right, those with core-dominant field topology. 
The results are represented at two initial magnetic field intensities at the polar surface), e.g., at $B=5 \times 10^{14}$ G (upper panels) and at $B=10^{13}$ G (bottom panels). }
\label{fig: comparison with light envelopes}
\end{figure*}

We now extend our analysis to accreted (light element) envelopes. The results of the comparison between light and heavy elements are shown in Fig. \ref{fig: comparison with light envelopes}. 
In the left panels, we display the results for models with crustal-confined magnetic fields and in the right panels those for core-dominant fields, for the two dipolar intensities
$B=5 \times 10^{14}$ G (upper panels) and at $B=10^{13}$ G (bottom panels). 

The same qualitative features discussed for iron envelopes are valid, but with luminosities shifted to slightly higher values (up to an order of magnitude) for accreted envelopes during the neutrino cooling epoch. Instead, it drops faster as soon as we enter in the photon cooling era. That is due to the even higher $T_s$ resulting from light elements in the envelope. A strong magnetic field enhances this effect. In the top-right panel of Fig. \ref{fig: comparison with light envelopes}, we clearly see how high luminosities ($> 10^{34}$ erg/s) are kept for about $10^4$ years, but then quickly drop below $10^{32}$ erg/s (and therefore objects become undetectable) after a few tens of kyr. Simulations using "old" non-magnetized models \citep{gudmundsson1983} do not allow to capture this behaviour.

\vspace{-6mm}
\section{Conclusions}
\label{sec: conclusions}

Understanding how young and middle-aged magnetars cool is of great importance for a correct interpretation of the observational data. In this work, we have revisited the cooling curves of NSs, focusing on the effect of the assumed envelope models (typically used as a boundary condition), and considering two extreme field topologies, crustal and core field. We noticed that the $T_b-T_s$ relation is very sensitive to the magnetic field strength. Although for relatively low magnetic fields, different magnetized iron envelopes predict similar effective surface temperatures. For relatively strong magnetic fields in the magnetar range there are substantial differences. For a given temperature at the base of the envelope ($T_b$), the most recent models (that incorporate better microphysics) predict surface temperatures ($T_s$) about a factor 2-3 higher than their predecessors. Since the flux scales as $T_s^4$, this correction significantly enlarges the photon luminosity, which in turn leads to a very fast transition from a high luminosity epoch (during the neutrino cooling era) to a very low luminosity (soon after we enter the photon cooling era).
This trend is very pronounced in models where the magnetic field threads the star's core and most electric currents circulate there. Conversely, in crustal-confined models, the additional energy released by Joule heating close to the star surface is very effective and governs the energy balance equation, which counterbalances the effect. 
Thus, depending on where the bulk of the electrical currents circulates, one can expect middle-aged magnetars which are relatively bright or sources with very low luminosities ($<10^{32}$ erg/s) which persistent emission is essentially undetectable as X-ray sources.
We stress again that a $10^4$ yr, high field NS with a core field (light and heavy element envelopes) can actually be much cooler than a similar NS with a pulsar-like field, only because of the effect of magnetic field in the envelope (see Fig.~\ref{fig: comparison with light envelopes} upper-right panel). This has potentially strong implications for population synthesis studies of the pulsar and magnetar populations because observational biases introduced by the lack of detectability of some class of sources affect the predictions of birth rates and field distributions. We plan to incorporate these effects in future works.

To briefly compare our results with observational data, one should only concentrate on objects with "Real ages" and that are at the extremes of our cooling curves: A) 1E 2259+586 (middle-aged magnetar) can only be explained with a crustal-field and magnetized light elements. B) All XDINS cannot be explained with core-fields. They necessarily need that the crustal-field has a strong component but the envelope can be light or heavy, magnetic or non-magnetic. C) CCOs are in an age and luminosity range that do not allow distinguishing between envelope models or magnetic topology. D) Middle-age faint pulsar such as PSR B2334+61 might be explained only with the fast decay of the light element envelope curves, since for low magnetic field NSs, light envelopes might produce cooler NSs than heavy elements for older ages ($>10^4$\,yr), regardless of the field configuration (see Fig.~\ref{fig: comparison with light envelopes} bottom panels). Ultimately, the existence of strongly magnetized neutron stars with detectable thermal emission at later times would be a strong argument in favor of a crustal magnetic field.

Our study highlights the importance of treating carefully all ingredients in the complex theory of NS cooling. Boundary conditions neglecting the role of the envelope, or using non-magnetized envelopes, can lead to discrepancies as large as one order of magnitude relative to observational data. On the other hand, an accurate estimation of surface luminosity is important to constrain any source property (e.g. surface B-field or age).

\vspace{-6mm}
\section*{Acknowledgements}

JAP acknowledges support from the Generalitat Valenciana grants PROMETEO/2019/071 and ASFAE/2022/026 (with funding from NextGenerationEU PRTR-C17.I1) and the AEI grant PID2021-127495NB-I00.
CD and NR are supported by the ERC Consolidator Grant “MAGNESIA” No. 817661 (PI: Rea) and this work has been carried out within the framework of the doctoral program in Physics of the Universitat Aut\`onoma de Barcelona and it is partially supported by the program Unidad de Excelencia Mar\'ia de Maeztu CEX2020-001058-M. 
DV is supported by the European Research Council (ERC) under the European Union’s Horizon 2020 research and innovation programme (ERC Starting Grant "IMAGINE" No. 948582, PI: DV).

\vspace{-6mm}

\section*{Data Availability}
Data available on request.

\vspace{-6mm}



\bibliographystyle{mnras}
\bibliography{mnras_template} 








\bsp	
\label{lastpage}
\end{document}